\documentclass[journal=jacsat,manuscript=article]{achemso}

\usepackage[version=3]{mhchem} 
 \SectionNumbersOn 
\usepackage{graphicx}
\usepackage{subfigure}
\usepackage{booktabs}
\usepackage{siunitx}
\DeclareSIUnit\angstrom{\protect \text {Å}}
\usepackage{comment} 

\usepackage{dcolumn}
\usepackage{enumitem}
\usepackage[final]{pdfpages}

\usepackage{doi}

\usepackage{bm}
\usepackage{mathrsfs}
\usepackage{amsmath}
\usepackage{mathtools}
\usepackage{physics}
\usepackage{braket}

\newcommand*\diff{\mathop{}\!\mathrm{d}}

\usepackage{caption}
\usepackage{subcaption}
\usepackage{tablefootnote}
\usepackage[font={footnotesize}]{caption}

\usepackage{hyperref}

\usepackage[]{xcolor}
\usepackage{soul}

\DeclareSIUnit{\atomicunit}{a.u.}
\DeclareSIUnit{\nanometer}{nm}

\author{Matteo Castagnola}
\affiliation{Department of Chemistry, Norwegian University of Science and Technology, 7491 Trondheim, Norway}

\author{Rosario R. Riso}
\affiliation{Department of Chemistry, Norwegian University of Science and Technology, 7491 Trondheim, Norway}

\author{Yassir El Moutaoukal}
\affiliation{Department of Chemistry, Norwegian University of Science and Technology, 7491 Trondheim, Norway}

\author{Enrico Ronca}
\affiliation{Dipartimento di Chimica, Biologia e Biotecnologie, Università degli Studi di Perugia, Via Elce di Sotto, 8,06123, Perugia, Italy}

\author{Henrik Koch}
\affiliation{Department of Chemistry, Norwegian University of Science and Technology, 7491 Trondheim, Norway}
\email{henrik.koch@ntnu.no}

\title {Strong coupling quantum electrodynamics Hartree-Fock response theory}

\begin{document}


\begin{tocentry}
\begin{center}
\includegraphics[width=\textwidth]{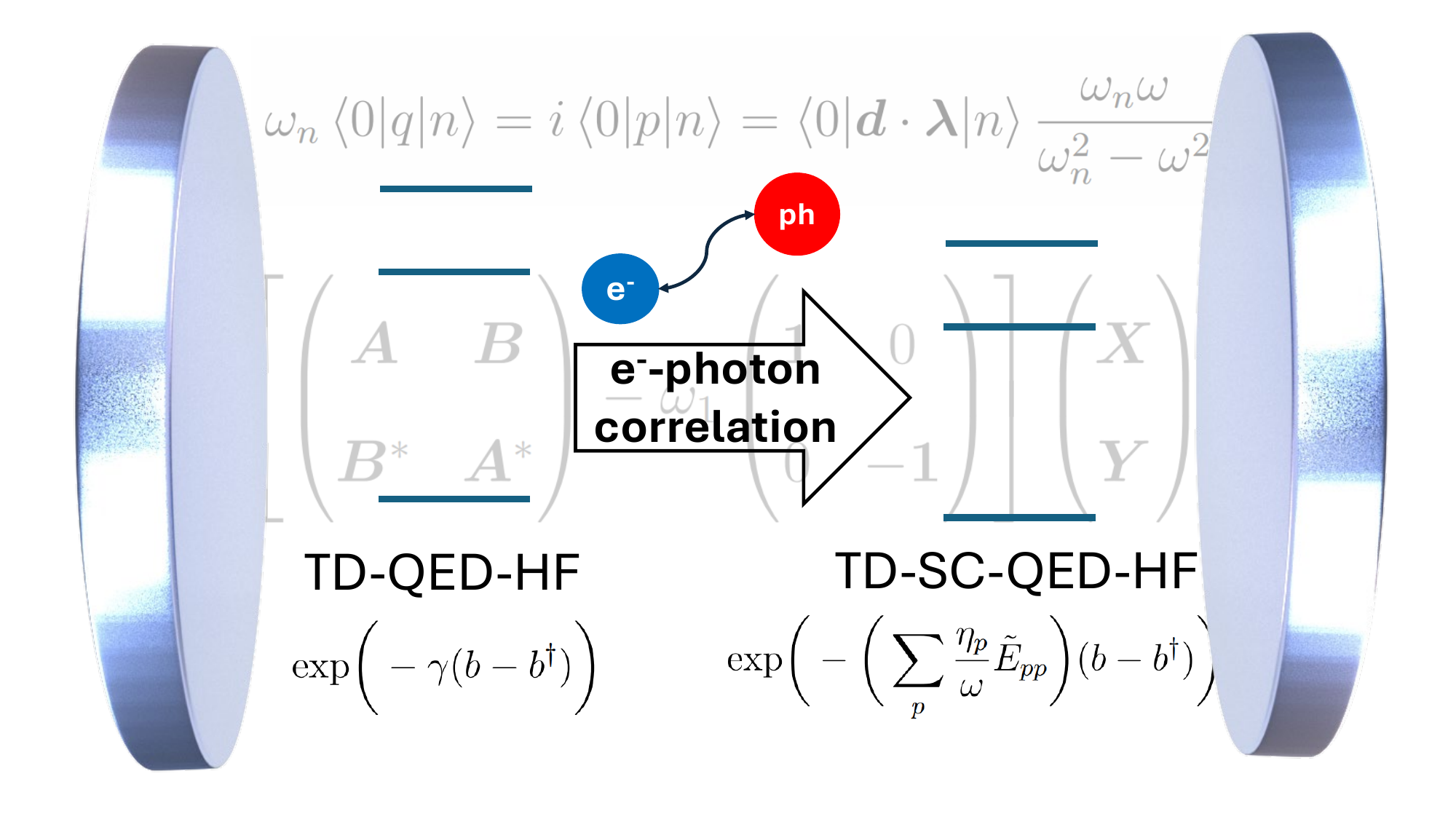}
\end{center}
\end{tocentry}







\begin{abstract}

The development of reliable \textit{ab initio} methods for light-matter strong coupling is necessary for a deeper understanding of molecular polaritons.
The recently developed strong coupling quantum electrodynamics Hartree-Fock model (SC-QED-HF) provides cavity-consistent molecular orbitals, overcoming several difficulties related to the simpler QED-HF wave function.
In this paper, we further develop this method by implementing the response theory for SC-QED-HF.
We compare the derived linear response equations with the time-dependent QED-HF theory and discuss the validity of equivalence relations connecting matter and electromagnetic observables.
Our results show that electron-photon correlation induces an excitation redshift compared to the time-dependent QED-HF energies, and we discuss the effect of the dipole self-energy on the ground and excited state properties with different basis sets.

\end{abstract}

\section{Introduction}

Molecular polaritons are hybrid light-matter states arising from the interaction of electronic (or vibrational) excitations with the electromagnetic modes of an optical resonator.\cite{ebbesen2023introduction, basov2020polariton}
Chemists are currently trying to exploit polaritons to control and alter chemical processes, such as ground state and photochemical reactions.\cite{Thomas2019, hutchison2012, zeng2023control, lee2024controlling, ahn2023modification}
The theoretical description of these systems, in which both the molecular properties and the photon components play a central role, has led to a merging of quantum optics models and quantum chemistry approaches.
The field of \textit{ab initio} quantum electrodynamics (QED) is rapidly growing and several methods have been proposed, such as quantum electrodynamics Hartree-Fock (QED-HF)\cite{haugland2020coupled}, polarized Fock states\cite{mandal2020polarized}, QED coupled cluster (QED-CC)\cite{haugland2020coupled, monzel2024diagrams, liebenthal2023assessing, mordovina2020polaritonic}, QED full configuration interaction (QED-FCI),\cite{haugland2020coupled} QED complete active space CI,\cite{vu2024cavity} and quantum electrodynamics density functional theory (QEDFT)\cite{ruggenthaler2014quantum, flick2015kohn}.
\textit{Ab initio} QED approaches describe the electronic structure (and the electron-photon coupling) at a high level of theory, although the computational complexity prevents the simulation of a large number of molecules.
Light-matter strong coupling has so far been achieved in the collective regime, i.e., with a relatively small coupling strength and several $\sim 10^5$ molecules coupled to the same electromagnetic mode of an optical resonator. 
Nevertheless, experiments are pushing toward larger coupling strengths (ultrastrong coupling regime).\cite{chikkaraddy2016single} 
The development of \textit{ab initio} QED methods is relevant as it allows for a nonperturbative description of light-matter coupling while simultaneously providing reliable modeling of the molecular electronic structure.
These methods also highlight subtleties in the description of the light-matter interplay,\cite{schafer2020relevance, rokaj2018light} thus providing a deeper understanding of the polaritonic wave functions.

In this paper, we develop and implement linear response equations for the strong coupling quantum electrodynamics Hartree-Fock method (SC-QED-HF).\cite{riso2022molecular, el2024toward}
The SC-QED-HF parametrization was introduced in 2022 by Riso et al.\cite{riso2022molecular} to solve issues in the Fock matrix of the QED-HF method, such as the origin-dependence of molecular orbitals for charged systems.
The SC-QED-HF model mixes the electronic and photonic degrees of freedom and becomes exact in the infinite coupling limit, introducing to some extent what we refer to as "electron-photon correlation".\cite{riso2022molecular, haugland2020coupled}
The SC-QED-HF wave function thus exhibits frequency dispersion (contrary to the QED-HF method), and the orbitals are consistent with increasing size of the system.\cite{moutaoukal2025strong, riso2022molecular}
A comparison of the QED-HF and SC-QED-HF results hence provides a simple way to assess the effect of electron-photon correlation on electronic and photonic properties.\cite{riso2022molecular}
A set of reliable molecular orbitals is also of great assistance in developing post-HF polaritonic methodologies since part of the electron-photon correlation is already accounted for by the orbitals, improving the convergence and the quality of the simulation.
Nevertheless, the changes in the electronic ground state induced by electronic strong coupling are usually small unless very large couplings are employed.
In contrast, polaritonic excitations have a significant photon component ($\sim 50\%$), and the Rabi splitting can reach a substantial fraction of the molecular excitation (ultrastrong coupling regime).
Therefore, the self-consistent feedback between the electronic and electromagnetic degrees of freedom can be expected to have a more relevant effect than for the ground state.
The developed response equations for SC-QED-HF then provide a straightforward way to assess the role of electron-photon correlation in the excited states and offer an additional step in developing consistent \textit{ab initio} methods for molecular polaritons.

The paper is organized as follows.
In \autoref{sec: theory}, we introduce the Pauli-Fierz Hamiltonian and develop the QED-HF and SC-QED-HF ground state and response equations.
We examine equivalence relations involving the transition moments for electronic and photonic observables and discuss the measure of the photon character of polaritonic excitations.
In \autoref{sec: results}, we analyze the simulations of excited states and transition properties obtained from the SC-QED-HF response equations, highlighting the effects of electron-photon correlation and dipole self-energy in the excited states.
Finally, in \autoref{sec: conclusions}, we summarize our findings and discuss future perspectives for \textit{ab initio} QED.

\section{Theory}\label{sec: theory}

The electromagnetic and electronic degrees of freedom are treated at the same level of theory using the single-mode Pauli-Fierz Hamiltonian in the dipole approximation, Born-Oppenheimer approximation, and length form\cite{ruggenthaler2023understanding, castagnola2024polaritonic, cohen2024photons}
\begin{align}\label{eq: dipole Hamiltonian}
    H&=\sum_{pq}h_{pq}E_{pq}+\frac{1}{2}\sum_{pqrs}g_{pqrs}e_{pqrs}+h_{nuc}\nonumber\\
    &+\frac{1}{2}\sum_{pqrs}(\bm{\lambda}\cdot \bm{d})_{pq}(\bm{\lambda}\cdot \bm{d})_{rs}E_{pq}E_{rs}\nonumber\\
    &-\sqrt{\frac{\omega}{2}}\sum_{pq}(\bm{\lambda}\cdot \bm{d})_{pq}E_{pq}(b^\dagger+b) \nonumber\\
    &+\omega {b}^\dagger{b},
\end{align}
where $b^\dagger$ ($b$) creates (annihilates) a photon of frequency $\omega$, $\bm{\lambda}=\lambda\bm{\epsilon}$ is the light-matter coupling strength along the field polarization $\bm{\epsilon}$, $E_{pq}$ and $e_{pqrs}$ are the spin adapted one- and two-electron operators in a molecular basis indexed with $p,q,r$, and $s$.\cite{helgaker2014molecular} 
The $\bm{d}$ operator is the dipole operator, and $h_{pq}$, $g_{pqrs}$ and $h_{nuc}$ are the one-electron integrals, two-electrons integrals, and nuclear repulsion, respectively.
The Hamiltonian in \autoref{eq: dipole Hamiltonian} includes the standard electronic Hamiltonian (first line of \autoref{eq: dipole Hamiltonian}), the energy of the photon field (last line), and the bilinear interaction (third line) between the molecular dipole and the photon field.
The second line of \autoref{eq: dipole Hamiltonian} is the dipole self-energy, which is necessary for the Hamiltonian to have a ground state.\cite{schafer2020relevance, rokaj2018light}

The Hamiltonian in \autoref{eq: dipole Hamiltonian} serves as the foundation for an \textit{ab initio} quantum electrodynamics treatment of the electron-photon system.
The wave function parametrization must then include parameters for both the electronic and electromagnetic degrees of freedom.
In the following section, we derive the ground state and response equations for QED-HF\cite{haugland2020coupled} and SC-QED-HF.\cite{riso2022molecular, el2024toward}

\subsection{QED-HF and SC-QED-HF ground state parametrization}

The QED-HF and SC-QED-HF wave functions are parametrized via a coherent state transformation $U_X$ (with $X = \text{QED-HF or SC}$) applied to the tensor product of a single Slater determinant $\ket{S}$ and the electromagnetic vacuum $\ket{0}$\cite{riso2022molecular, haugland2020coupled}
\begin{equation}
    \ket{\text G\text S} = U_X \ket{\text S} \otimes \ket{0}.
\end{equation}
The state transformations $U_X$ are given by
\begin{align}
    U_{\text{QED-HF}} &= \text{exp}\bigg( -\gamma (b-b^\dagger) \bigg) \label{eq: U QED HF}\\ 
    U_{\text{SC}} &= \text{exp}\bigg( -\bigg(\sum_{p}\frac{\eta_p}{\omega}\tilde{E}_{pp}\bigg) (b-b^\dagger) \bigg). \label{eq: U SC}
\end{align}
In \autoref{eq: U SC}, the parameters $\eta_p$ are orbital-specific coherent state parameters associated with the orbital basis that diagonalizes the dipole interaction operator, indicated with a $\sim$
\begin{equation}
    (\bm d \cdot \bm \lambda) = \sum_p \widetilde{(\bm d \cdot \bm \lambda)}_{pp} \tilde E_{pp}.\label{eq: dipole basis}
\end{equation}
The orbital and photon parameters are obtained, using the variational principle, by minimizing the expectation value of the Hamiltonian in \autoref{eq: dipole Hamiltonian}.
In the following, the tilde spin-adapted operators $\tilde E_{pq}$ and $\tilde e_{pqrs}$ will refer to the dipole basis of \autoref{eq: dipole basis}. \\

The optimal photon parameter $\gamma$ of QED-HF in \autoref{eq: U QED HF} is\cite{haugland2020coupled}
\begin{equation}
    \gamma = -\frac{\bm \lambda \cdot \braket{\bm d}_{\text{QED-HF}}}{\sqrt{2 \omega}}, \label{qe: gamma QEDHF}
\end{equation}
where $\braket{\bm d}_{\text{QED-HF}}$ is the expectation value of the dipole operator for the optimal QED-HF Slater determinant. 
Transforming the Hamiltonian in \autoref{eq: dipole Hamiltonian} with the QED-HF operator $U_{\text{QED-HF}}$, we obtain
\begin{align}
    U^\dagger_{\text{QED-HF}}H\,U_{\text{QED-HF}} =&\sum_{pq}h_{pq}E_{pq}+\frac{1}{2}\sum_{pqrs}g_{pqrs}e_{pqrs}+h_{nuc}  \nonumber \\
    & + \frac{1}{2}\sum_{pqrs} (\bm\lambda \cdot(\bm{d}-\braket{\bm{d}}_{\text{QED-HF}}) )_{pq} (\bm\lambda \cdot(\bm{d}-\braket{\bm{d}}_{\text{QED-HF}}) )_{rs} E_{pq} E_{rs} \nonumber \\
    & - \sqrt{\frac{\omega}{2}} \sum_{pq}(\bm{\lambda}\cdot (\bm{d}-\braket{\bm{d}}_{\text{QED-HF}}) )_{pq}E_{pq} (b+b^\dagger) + \omega b^\dagger b, \label{eq: Ham coherenst transf}
\end{align}
which is manifestly origin invariant even for charged systems.
The molecular orbitals are obtained from the QED Fock matrix\cite{haugland2020coupled}
\begin{align}
    &F_{pq}=F_{pq}^e+\frac{1}{2}\bigg(\sum_a(\bm{\lambda}\cdot\bm{d}_{pa})(\bm{\lambda}\cdot\bm{d}_{aq})-\sum_i (\bm{\lambda}\cdot\bm{d}_{pi})(\bm{\lambda}\cdot\bm{d}_{iq})\bigg), \label{eq: Fock QEDHF}
\end{align}
where $F_{pq}^e$ is the standard electronic Fock matrix.\cite{helgaker2014molecular}
In \autoref{eq: Fock QEDHF}, indexes $i$ and $a$ refer to occupied and virtual orbitals, respectively.
The occupied and virtual blocks of the Fock matrix in \autoref{eq: Fock QEDHF} are origin-dependent for charged systems.
Therefore, the molecular orbitals change accordingly (although the total energy is origin invariant).

The SC-QED-HF parametrization in \autoref{eq: U SC} is introduced to obtain more consistent molecular orbitals.\cite{riso2022molecular, el2024toward}
The transformation $U_{\text{SC}}$ explicitly correlates electronic and photonic degrees of freedom: the electronic creation operators $\tilde a^\dagger_p$, when transformed with $U_{\text{SC}}$, reads
\begin{equation}
    U_{\text{SC}}^\dagger \;\tilde a_p^\dagger \;U_{\text{SC}} = \tilde a_p^\dagger \;\text{exp}\bigg({\frac{\eta_p}{\omega}(b-b^\dagger)}\bigg).\label{eq: el-ph mix}
\end{equation}
That is, each electronic creation (annihilation) operator in the dipole basis is dressed with a coherent state of the photon field, and the SC-transformed Hamiltonian reads
\begin{equation}\label{eq: SC Hamiltoninan}
    \begin{split}
    {H}_{\mathrm{SC}} = U^\dagger_{\mathrm{SC}} H U_{\mathrm{SC}} & = \sum_{pq} \tilde{h}_{pq} \tilde{E}_{pq}\exp (\frac{1}{\omega}(\eta_p-\eta_q)(b-b^\dagger)) \\
    &+ \frac{1}{2}\sum_{pqrs}\tilde{g}_{pqrs}\tilde{e}_{pqrs}\exp(\frac{1}{\omega}(\eta_p+\eta_r-\eta_q-\eta_s)(b-b^\dagger)) + h_{nuc}\\
    & + \omega \Big(b^\dagger-\sum_p\bigg(\frac{\lambda}{\sqrt{2\omega}}\widetilde{(\bm{d}\cdot\bm{\epsilon})}_{pp} - \frac{\eta_p}{\omega}\bigg)\tilde{E}_{pp}\Big)\Big(b-\sum_p\bigg(\frac{\lambda}{\sqrt{2\omega}}\widetilde{(\bm{d}\cdot\bm{\epsilon})}_{pp} - \frac{\eta_p}{\omega}\bigg)\tilde{E}_{pp}\Big) .
    \end{split}
\end{equation}
The electromagnetic and electronic degrees of freedom are thus entangled via the transformation in \autoref{eq: U SC}, and simultaneous optimization of the $\eta_p$ and orbital parameters is required.\cite{el2024toward}
We notice that, from \autoref{eq: U SC}, we can recover the QED-HF parametrization by setting $\eta_p = \omega\gamma/N$, where $N$ is the number of electrons in the system, and therefore the SC-QED-HF variational energy is always lower than the QED-HF energy.

The SC-QED-HF model provides a significant improvement over the simpler QED-HF wave function.
First, the SC-QED-HF parametrization solves the origin dependence issues for charged systems.
Second, the wave function shows dispersive behavior with the cavity frequency, contrary to the QED-HF energy and orbitals, which are $\omega$-independent.\cite{riso2022molecular, el2024toward}
Moreover, the QED-HF orbitals show a troublesome behavior when two subsystems are separated over large distances, which can be particularly problematic when addressing perturbation theory or the collective coupling regime.\cite{moutaoukal2025strong}
A set of QED-consistent molecular orbitals is also relevant for the development of reliable post-HF methods.

\subsubsection{Infinite coupling limit and electron-photon correlation}

The SC-QED-HF parametrization is introduced from the infinite coupling limit ${H}_\infty$ of the dipole Hamiltonian\cite{riso2022molecular, ashida2021cavity, di2019resolution}
\begin{equation}
    {H}_\infty = \omega b^\dagger b - \lambda \sqrt{\frac{\omega}{2}} ( {\bm{d}}\cdot \bm{\epsilon}) ( b^{\dagger}+b ) + \frac{\lambda^2}{2}  ( {\bm{d}} \cdot \bm{\epsilon})^2 .\label{eq: H infty}
\end{equation}
If we set $\eta_p =  \sqrt{\frac{\omega}{2}} \widetilde{(\bm{d}\cdot\bm{\lambda})}_{pp}$ in $U_{\text{SC}}$ and transform $H_{\infty}$, we find that the transformation diagonalizes the Hamiltonian in \autoref{eq: H infty}.
Therefore, as the coupling increases, the SC-QED-HF model approaches the exact solution in which the electron and photon degrees of freedom are deeply entangled (see \autoref{eq: el-ph mix}).\cite{riso2022molecular, ashida2021cavity, di2019resolution}
On the other hand, the QED-HF orbitals also approach the dipole basis, but $U_{\text{QED-HF}}$ fails to properly correlate the electrons and the photon field.

In the HF ansatz, the electrons are treated independently from one another (except for the Fermi correlation arising from the wave function antisymmetry).
The instantaneous electronic position is thus not relevant since each electron perceives a mean-field Coulomb effect from the electrons in the other occupied orbitals.
Therefore, the HF method is considered uncorrelated.
In electronic structure theory, electron-electron correlation is then defined as the difference between the full configuration-interaction (FCI) energy and the HF energy, within the employed basis set.\cite{helgaker2014molecular}
The same definition is used for the correlation energy in approximate methods such as coupled cluster or truncated CI.
In the \textit{ab initio} QED framework, we introduce the photons as additional boson particles.
Therefore, electron correlation is modified, e.g., by additional photon-mediated electron interactions (electron-photon-electron interactions).
The entanglement of the electromagnetic and electronic degrees of freedom requires a definition of electron-photon correlation.
Since the electronic and electromagnetic degrees of freedom are deeply intertwined, it can be cumbersome to separate electronic and electron-photon effects, especially for large $\lambda$ and highly correlated methods.
However, SC-QED-HF becomes exact in the infinite coupling limit (where electron-photon correlation dominates over any electronic effect), and the comparison between the mean field QED-HF and the entangled SC-QED-HF wave function provides a simple measure of the electron-photon correlation in the ground state.
For the excited states, it is even harder to provide an effective definition of electron-photon correlation since matter and photon excitations have a similar weight in the polaritonic states.
We will nevertheless argue that SC-QED-HF response theory can lead to a simple understanding of excited state electron-photon correlation.

\subsection{Response theory for SC-QED-HF}

The QED-HF and SC-QED-HF response theory is based on the TD-QED-HF parametrization\cite{castagnola2024polaritonic}
\begin{align}
    \ket{\text R(t)}&=\text{exp}\big(-i\Lambda(t)\big)\ket{\text{S}, 0}\nonumber\\
    &=\text{exp}\big(-i\bm{\kappa}(t)\big)\text{exp}\bigg(-i\big(\gamma(t)\; b^\dagger +\gamma^*(t) \;b\big)\bigg)\ket{\text{R}}, \label{eq: tdqedhfpar}
\end{align}
where $\ket{\text{R}}=\ket{\text{S}}\otimes\ket{0}$ is the optimal determinant in the electromagnetic vacuum, $\bm{\kappa}(t)$ is an orbital rotation operator
\begin{equation}
    \bm{\kappa}(t)=\frac{1}{\sqrt{2}}\sum_{ai}\big(\kappa_{ai}E_{ai}+\kappa^*_{ai}E_{ia}\big)
\end{equation}
where $i$ ($a$) label occupied (virtual) orbitals of the reference determinant, and $\gamma$ describes the field evolution.
In \autoref{eq: tdqedhfpar}, $\ket{\text{S}}$ is the reference (SC-)QED-HF Slater determinant $\ket{\text{(SC-)QED-HF}}$, and the parametrization refers to the transformed Hamiltonian 
\begin{equation}
    H_X=U^\dagger_X\,H\, U_X,    \label{eq: H transf X}
\end{equation}
with $X = \text{QED-HF or SC}$.
That is, the TD-QED-HF and TD-SC-QED-HF wave functions refer respectively to the Hamiltonians in \autoref{eq: Ham coherenst transf} and \autoref{eq: SC Hamiltoninan}.
The transformation $U_{\text{SC}}$ ensures that for the infinite coupling limit in \autoref{eq: H infty}, the parametrization in \autoref{eq: tdqedhfpar} recovers the exact solutions.
The response equations for the Hamiltonian $U^\dagger_X\,(H+V)\, U_X$, where $ V$ is an external perturbation operator, are obtained from a perturbation expansion in the frequency domain\cite{olsen1985linear}
\begin{align}
    \Lambda(t) &= \Lambda^{(1)}(t) + \Lambda^{(2)}(t)+ \dots \nonumber \\
    &= \int \diff\omega_1\;e^{-i\omega_1 t}\Lambda^{\omega_1} + \frac{1}{2} \int \diff\omega_1\,\diff\omega_2\;e^{-i\omega_1 t}e^{-i\omega_2 t}\Lambda^{\omega_1,\omega_2} + \dots\label{eq: per exp t op}
\end{align}
where
\begin{align}
    \Lambda^{\omega_1} &=\frac{1}{\sqrt{2}}\sum_{ai}\big(\kappa_{ai}^{\omega_1}E_{ai}+[\kappa_{ai}^{-{\omega_1}}]^*\;E_{ia}\big)+\big(\gamma^{{\omega_1}} \;b^\dagger +[\gamma^{-{\omega_1}}]^* \;b\big)\label{eq: omega-qed-hf-par}\\
    \Lambda^{\omega_1,\omega_2} &=\frac{1}{\sqrt{2}}\sum_{ai}\big(\kappa_{ai}^{\omega_1,\omega_2}E_{ai}+[\kappa_{ai}^{-\omega_1,-\omega_2}]^*\;E_{ia}\big)+\big(\gamma^{\omega_1,\omega_2} \;b^\dagger +[\gamma^{-\omega_1,-\omega_2}]^* \;b\big).
\end{align}
Using \autoref{eq: per exp t op}, we define the response functions from the expectation value of an operator $A$
\begin{align}
    \braket{A}(t)&= \braket{A}_\text{R} + \int \diff\omega_1\;e^{-i\omega_1 t}\braket{\braket{A;V^{\omega_1}}}_{\omega_1} \nonumber\\
    &+ \frac{1}{2} \int \diff\omega_1\,\diff\omega_2\;e^{-i\omega_1 t}e^{-i\omega_2 t}\braket{\braket{A;V^{\omega_1},V^{\omega_2}}}_{\omega_1,\omega_2} + \dots 
\end{align}

The linear response equations, which can be obtained following the same derivation as Olsen and J{\o}rgensen based on Eherenfest's theorem,\cite{olsen1985linear} are the same for the two parametrizations
\begin{equation}\label{eq: tdhf_response}
    \left[\begin{pmatrix}
     \bm{A}&\bm{B}\\
     \bm{B}^*&\bm{A}^*
    \end{pmatrix}-\omega_1 \begin{pmatrix}
     \bm 1&0\\
     0&-\bm 1
    \end{pmatrix}\right]\begin{pmatrix}
     \bm{X}\\\bm{Y}
    \end{pmatrix}=i\begin{pmatrix}
     \bm{g}_1\\ \bm{g}_2
    \end{pmatrix}
    \equiv i\bm{g},
\end{equation}
where the vectors $\bm{X}$ and $\bm{Y}$ collect the Fourier transformed parameters
\begin{equation}
    \bm{X}=\begin{pmatrix}
    \gamma^{\omega_1}\\
    \kappa_{ai}^{\omega_1}
    \end{pmatrix}\quad \bm{Y}=\begin{pmatrix}
    [\gamma^{-{\omega_1}}]^*\\
    [\kappa_{ai}^{-{\omega_1}}]^*
    \end{pmatrix}.
\end{equation}
The right hand side of Eq.~\eqref{eq: tdhf_response} is the generalized gradient
\begin{equation}\label{eq:generalized_gradient}
    \bm{g}_1=\begin{pmatrix}
    \braket{[b,V_X^{\omega_1}]}_\mathrm{R}\\
    \frac{1}{\sqrt{2}}\braket{[E_{ia},V_X^{\omega_1}]}_\mathrm{R}
    \end{pmatrix}\quad \bm{g}_2=\begin{pmatrix}
    \braket{[b^\dagger,V_X^{\omega_1}]}_\mathrm{R}\\
    \frac{1}{\sqrt{2}}\braket{[E_{ai},V_X^{\omega_1}]}_\mathrm{R}
    \end{pmatrix},
\end{equation}
and the explicit expressions of $\bm{A}$ and $\bm{B}$ are
\begin{align}
    \bm{A}&=\begin{pmatrix}
     \braket{[b,[H_X,b^\dagger]]}_\text{R}&\frac{1}{\sqrt{2}}\braket{[b,[H_X,E_{ai}]]}_\text{R}\\
     \frac{1}{\sqrt{2}}\braket{[b^\dagger,[H_X,E_{ia}]]}_\text{R}&\frac{1}{2}\braket{[E_{jb},[H_X,E_{ai}]]}_\text{R}
    \end{pmatrix}\label{eq: Asingle} \\
    \bm{B}&=\begin{pmatrix}
     \braket{[b,[H_X,b]]}_\text{R}&\frac{1}{\sqrt{2}}\braket{[b,[H_X,E_{ia}]]}_\text{R}\\
     \frac{1}{\sqrt{2}}\braket{[b^\dagger,[H_X,E_{ai}]]}_\text{R}&\frac{1}{2}\braket{[E_{bj},[H_X,E_{ai}]]}_\text{R}\label{eq: Bsingle}
    \end{pmatrix}.
\end{align}
Therefore, \autoref{eq: tdhf_response} is a generalization of the Casida equations of TDHF in molecular response theory, including additional parameters describing the photon field.\cite{flick2019light, castagnola2024polaritonic, ruggenthaler2018quantum, casida2009time} 
The difference between the QED-HF and SC-QED-HF response equations is embedded in the picture change $U_X$ and the orbitals of the reference determinant $\ket{\text{S}}$.
Since the SC-QED-HF model introduces electron-photon correlation via the $U_{\text{SC}}$ transformation, a comparison between TD-QED-HF and TD-SC-QED-HF reveals the impact of electron-photon correlation on the excited states.

\subsubsection{Equivalence relations}

The exact linear response functions fulfill the equation of motion\cite{olsen1985linear}
    \begin{align}
    \omega_1\braket{\braket{A;V^{\omega_1}}}_{\omega_1}&=\braket{\braket{[A,H];V^{\omega_1}}}_{\omega_1}+\braket{0|[A,V^{\omega_1}]|0}\label{eq: fir_ord EOM respo fun}. 
\end{align}
From the position-momentum relation
\begin{equation}\label{eq: pr comm ham}
    i \bm p_i = [\bm r_i,H],
\end{equation}
which also holds for the dipole Hamiltonian in length form in \autoref{eq: dipole Hamiltonian}, we obtain the equivalence between the dipole and velocity formulation of the transition moments
\begin{equation}\label{eq: dipole-velocity-form}
    \omega_n \braket{0|\bm{r}_i|n}=i\braket{0|\bm{p}_i|n},
\end{equation}
where $\omega_n$ is the excitation energy from the ground state to the excited state $\ket{n}$.

Analogous equivalence relations can be derived for photon observables.\cite{castagnola2024polaritonic}
In fact, an equation analogous to \autoref{eq: pr comm ham} also holds for the photon coordinate $q= \frac{1}{\sqrt{2\omega}}(b+b^\dagger)$ and photon momentum $p=i\sqrt{\frac{\omega}{2}}(b^\dagger-b)$ for the Hamiltonian in \autoref{eq: dipole Hamiltonian}
\begin{equation}\label{eq: qp comm ham}
    i  p = [q, H],
\end{equation}
which provides a relation between the corresponding transition moments
\begin{equation}\label{eq: dipole-velocity-form photons}
    \omega_n \braket{0|q|n}=i\braket{0|p|n}.
\end{equation}
It is also possible to derive relations that connect photon and electronic quantities.
For the operators $b$ and  $b^\dagger$, we can obtain the identities
\begin{align}\label{eq: photon_eq_ref}
\omega_1\braket{\braket{b;V^{\omega_1}}}_{\omega_1}&=\langle{\langle{-\sqrt{\frac{\omega}{2}}\bm{d}\cdot\bm{\lambda}
-\omega b;V^{\omega_1}}}\rangle\rangle_{\omega_1}+\braket{0|[b,V^{\omega_1}]|0}\\
\omega_1\braket{\braket{b^\dagger;V^{\omega_1}}}_{\omega_1}&=\langle{\langle{\sqrt{\frac{\omega}{2}}\bm{d}\cdot\bm{\lambda}
+\omega b^\dagger;V^{\omega_1}}}\rangle\rangle_{\omega_1}+\braket{0|[b^\dagger,V^{\omega_1}]|0},
\end{align}
from which we obtain\cite{castagnola2024polaritonic}
\begin{align}
    \omega_n\braket{0|q|n}&=i\braket{0|p|n}=\braket{0|\bm{d}\cdot\bm{\lambda}|n}\frac{\omega_n\omega}{\omega_n^2-\omega^2}.\label{eq: photo_momenta_conjugate}
\end{align}
The relation in \autoref{eq: photo_momenta_conjugate} relates electromagnetic and molecular observables, reflecting the intrinsic connection of the electronic and photonic degrees of freedom.

These equivalence relations can have relevant physical meanings and can be of practical importance in simulations.
For instance, from the quadratic response function, it is possible to show that $\braket{n|\hat{E}_\perp|n}=0$, where $\hat{E}_\perp=-\bm \lambda\big(\bm \lambda \cdot \bm d\big)-\bm\lambda\omega q$ is the electric field operator, which guarantees that the ground state is non-radiating.
At the same time, \autoref{eq: dipole-velocity-form} is fundamental for obtaining origin invariant optical rotational strengths\cite{pedersen2004origin}.
Although relevant, these equivalence relations are not guaranteed for approximate wave functions described in a finite basis set.
In particular, the relation in \autoref{eq: pr comm ham} requires basis set completeness,\cite{helgaker2014molecular} while \autoref{eq: qp comm ham} does not depend on the basis set size.
In Ref.\citenum{castagnola2024polaritonic}$ $, it was shown that the relations \autoref{eq: dipole-velocity-form} and \autoref{eq: photo_momenta_conjugate} hold for TD-QED-HF.
In particular, while \autoref{eq: dipole-velocity-form} holds only for a complete basis set, \autoref{eq: photo_momenta_conjugate} if fulfilled independently of basis set size\footnote{the numerical values of the transition moments change with the basis set, but for each basis set the relation \autoref{eq: photo_momenta_conjugate} is fullfilled}. 
For the SC-QED-HF wave function, the $U_{\text{SC}}$ transformation mixes electronic and photonic operators
\begin{align}
    U_{\text{SC}}^\dagger \, \tilde E_{pq} \, U_{\text{SC}} &= \tilde E_{pq} \;\text{exp}\bigg({\frac{\omega_p-\omega_q}{\omega}(b-b^\dagger)}\bigg) \\
    U_{\text{SC}}^\dagger  \,b\,U_{\text{SC}} &= b + \sum_p\frac{\eta_p}{\omega}\tilde E_{pp},
\end{align}
which complicates the analysis.
However, since the transformation $U_{\text{SC}}$ commutes with the photon momentum $p$ and the dipole operator $\bm d$, it is possible to show that the relations in \autoref{eq: dipole-velocity-form}  (for a complete basis set) and \autoref{eq: photo_momenta_conjugate} are fulfilled for the proposed TD-SC-QED-HF model (see the Supporting Information).

\subsubsection{Photon character and picture change}

Polaritons are states of hybrid light-matter character, and the photon component plays a relevant role in their properties.
It is then interesting to discuss the relative contribution of the photon and matter excitations in the polaritonic wave function.
In this section, we highlight the difficulties encountered in providing a clear definition of the "photon character" of excitations.

The response parameter $\gamma$, associated with the photon operator $b^\dagger$ in \autoref{eq: omega-qed-hf-par}, can provide a measure of the photon weight $\vartheta_n$ in the excited state $\ket{n}$\cite{yang2021quantum}
\begin{equation}\label{eq: photon char param}
    \vartheta_n = \gamma_n^2.
\end{equation}
Nevertheless, care is advised in adopting such an interpretation since the analysis is hindered by the picture change $U_X$ in \autoref{eq: H transf X}, especially for SC-QED-HF as $U_{\text{SC}}$ mixes electronic and photonic components.
Moreover, the Hamiltonian in \autoref{eq: dipole Hamiltonian} originates from the Power–Zienau–Woolley transformation of the minimal coupling Hamiltonian (within the dipole approximation).\cite{ruggenthaler2023understanding, cohen2024photons}
As a consequence, the photon coordinate $q = \frac{1}{\sqrt{2\omega}}(b+b^\dagger)$ is connected to the displacement field $\bm{D}$ of the macroscopic Maxwell's equations rather than the microscopic electric field $\bm{E}$.\cite{ruggenthaler2023understanding, castagnola2024polaritonic, schafer2020relevance, cohen2024photons}
Since the electronic and electromagnetic degrees of freedom are intertwined, it is then questionable to think of $b^\dagger$ as a "purely photonic" operator.
In addition, the photon component of the ground state is nonnegligible since we describe the electron-photon interaction nonperturbatively within an \textit{ab initio} framework.
Our ground state results thus differ from the simplified Jaynes-Cummings model,\cite{jaynes1963comparison} where the rotating wave approximation and the neglect of the dipole self-energy results in an unmodified electronic ground state in the electromagnetic vacuum.
It is accordingly difficult to provide a clear definition of the "photon character" of excitations, and interpretations of such quantities can be misleading, especially for large coupling strengths $\lambda$.

Photonic observables can provide a more rigorous measure of the photon role in the wave function.
The photon count $ b^\dagger b$ is a natural choice, though its representation changes with $U_X$ and it is thus generally nonzero for the ground state.
Moreover, the photon number operator $ b^\dagger b$ for the length Hamiltonian does not correspond to the photon number operator in the velocity form.
Computing $\braket{n|b^\dagger b|n}$ (or the corresponding quantities for the SC and velocity representations) further requires the quadratic response functions.\cite{olsen1985linear}
Therefore, we suggest a different quantity as a measure of the photon character of a polaritonic excitation that can be complementary to \autoref{eq: photon char param}.
The photon momentum $i p = \sqrt{\frac{\omega}{2}}(b-b^\dagger)$ commutes with the operator $U_X$ and is connected to the vector potential operator in the velocity representation.
The expectation value $\braket{n|p|n}$ (which would, in principle, require quadratic response functions) is identically zero for any real wave function.
Using the operator $p$ is thus convenient since its expectation value vanishes identically for the (SC-)QED-HF ground state and has a clear physical interpretation.
Since $\braket{n|p|n} = 0 $, we rely on the transition moments
\begin{equation}
    \abs{\braket{\text{R}|p|n}}^2,
\end{equation}
which can be computed from the linear response equations, as an indication of how the photon component changes following a transition from the ground state $\ket{\text{R}}$ to the excited state $\ket{n}$.

\section{Results}\label{sec: results}

In this section, we present the results for the TD-SC-QED-HF linear response method outlined in the previous section, also providing a comparison with the TD-QED-HF model.
The QED-HF, SC-QED-HF, and TD-QED-HF equations are implemented in the development branch of the $e^\mathcal{T}$ program, an open-source electronic structure program.\cite{folkestad2020t}
The TD-SC-QED-HF model has been implemented in a local branch of the $e^\mathcal{T}$ program.

\subsection{Equivalence relations and dipole self-energy}
In \autoref{tab: LP transition moments}, we report transition observables computed using different basis sets for the lower polariton (LP) of a formaldehyde molecule in an optical cavity with field polarization parallel to the transition dipole of the first bright molecular excitation.
The oscillator strengths in velocity $f_v$ and length $f_l$ forms
\begin{equation}
    f_l = \frac{2}{3} \omega_{LP} \, \abs{\braket{\text{GS}|\bm{d}|\text{LP}}}^2 \qquad f_v = \frac{2}{3} \frac{1}{\omega_{LP}} \abs{\braket{\text{GS}|\bm{p}|\text{LP}}}^2,
\end{equation}
where $\omega_{LP}$ is the LP excitation energy, are connected to the intensity of the electronic excitation.
\begin{table}
    \centering
    \caption{Oscillator strengths (in length gauge $f_l = \frac{2}{3} \omega_{LP} \, \abs{\braket{\text{GS}|\bm{d}|\text{LP}}}^2$ and velocity gauge $f_v = \frac{2}{3} \frac{1}{\omega_{LP}} \abs{\braket{\text{GS}|\bm{p}|\text{LP}}}^2$) and transition moments for the photon coordinate $q=\frac{1}{\sqrt{2\omega}}(b+b^\dagger)$ and photon momentum $ip =\sqrt{\frac{\omega}{2}}(b-b^\dagger)$ for the first polaritonic state (LP) of formaldehyde computed using TD-SC-QED-HF in different basis sets. 
    The photon frequency is set to $\omega =$ \qtylist{0.311860}{\atomicunit}, the coupling strength is $\lambda =$ \qtylist{0.01}{\atomicunit}, and the molecular geometry is reported in the Supporting Information.
    The photon field is perpendicular to the C-O bond in the molecular plane.
    In the table, we also report the photon and electron-photon quantities shown in \autoref{eq: photo_momenta_conjugate}. 
    While the position-momentum equivalence in \autoref{eq: dipole-velocity-form} is only fulfilled for a complete basis set, the relation in \autoref{eq: photo_momenta_conjugate} is independent of the basis set choice.
    }
    \begin{tabular}{|c|c|c|c|c|c|c|}
         \hline basis &   $f_l$  &   $f_v$  &    $q_{0n}$ [a.u.]&   $ip_{0n}$ [a.u.]&   $\omega_nq_{0n}$ [a.u.]  &   $\frac{\omega\omega_n(\bm\lambda\cdot\bm d)_{0n}}{\omega_n^2-\omega_2}$ [a.u.]\\ \hline
         sto-3g &   0.0001  &  0.0001  &  1.26635123  &  0.39477749 &   0.39477749 &   0.39477749\\
6-31g  &  0.0014  &  0.0011  &  1.26577915  &  0.39432973  &  0.39432973  &  0.39432973\\
6-31g*   & 0.0015 &   0.0013 &   1.26582131  &  0.39432455  &  0.39432455  &  0.39432455\\
6-31g**  &  0.0016  &  0.0014   & 1.26580336   & 0.39431348 &   0.39431348  &  0.39431348\\
6-31+g**  &  0.0042   & 0.0041  &  1.25502739   & 0.39067397  &  0.39067397  &  0.39067397\\
6-31++g**  &  0.0151   & 0.0140  &  1.15402184 &   0.35853984  &  0.35853984  &  0.35853984\\
aug-cc-pvdz   & 0.0222  & 0.0219   & 1.01273656 &   0.31418802   & 0.31418802&    0.31418802\\
d-aug-cc-pvdz &  0.0258  &  0.0252 &   0.88685195  &  0.27480100  &  0.27480100   & 0.27480100\\
d-aug-cc-pvtz   & 0.0203   & 0.0202  & 1.03445988  &  0.32103706   & 0.32103706  &  0.32103706\\\hline
    \end{tabular}
    \label{tab: LP transition moments}
\end{table}
Based on the equivalence relation in \autoref{eq: dipole-velocity-form}, the velocity and length gauge oscillator strengths should converge when approaching basis set completeness.
In \autoref{tab: LP transition moments}, we also report the transition moments for the photon coordinate $q$ and momentum $p$, and focus on the equivalence relation in \autoref{eq: photo_momenta_conjugate}.
While the commutator relation in \autoref{eq: pr comm ham} requires a complete basis set to be fulfilled,\cite{helgaker2014molecular} \autoref{eq: qp comm ham} is basis-set independent.
Since the optimization of the SC-QED-HF wave function is not performed within a truncated photon subspace, the photonic equivalence relations in \autoref{eq: photo_momenta_conjugate} are fulfilled exactly for any basis set size for TD-SC-QED-HF, similarly to the TD-QED-HF model.\cite{castagnola2024polaritonic}
If we rely on the Tamm-Dancoff approximation (TDA) by disregarding the $\bm B$ block in the response equations in \autoref{eq: tdhf_response} (which is equivalent to a CIS model), the relations in \autoref{eq: dipole-velocity-form} and \autoref{eq: dipole-velocity-form photons} are not guaranteed anymore.

For the calculations reported in \autoref{tab: LP transition moments}, we used the Hamiltonian in \autoref{eq: dipole Hamiltonian}, which includes the dipole self-energy (DSE).
The DSE ensures the Hamiltonian to be bounded from below, and thus, the system has a stable ground state.\cite{rokaj2018light, schafer2020relevance}
If the DSE is disregarded, the energy of the system can become infinitely low by displacing the electrons far away from the nuclear core along the polarization direction.
However, since the atomic orbitals are centered on the nuclei, such electronic displacement is generally hard to produce within a finite basis set.
It is thus worth investigating how the molecular properties behave when the basis set is increased without having the DSE in the Hamiltonian.
In \autoref{tab: LP nodse sc}, we report the ground state energy, the LP excitation energy, and transition properties for a formaldehyde molecule, computed including or disregarding the DSE in the Pauli-Fierz Hamiltonian.
The difference in the computed ground state energy is of the order of $\sim 10^{-4}$ a.u., which, for $\lambda =$ \qtylist{0.01}{\atomicunit}, is consistent with a perturbative analysis of the DSE.
However, we see that the energy difference does increase with the basis set size.
\begin{table}
    \centering
    \caption{Ground state (GS) energy, lower polariton (LP) excitation energy, length-gauge oscillator strength $f_l$ and photon transition moments $q_{0n}$ and $ip_{0n}$ for a formaldehyde molecule in different basis set with and without the dipole self-energy (DSE) term in the Hamiltonian.
    The photon field is aligned perpendicular to the C-O bond in the molecular plane.
    The photon frequency is set to $\omega =$ \qtylist{0.311860}{\atomicunit}, the coupling strength is $\lambda =$ \qtylist{0.01}{\atomicunit}, and the molecular geometry is reported in the Supporting Information.}
    \begin{tabular}{|c|c|c|c|c|c|}
         \hline basis &   E$_\text{GS}$  [a.u.]& $\omega_{LP} $ [a.u.] &   $f_l$  &    $q_{0n}$ [a.u.]&   $ip_{0n}$ [a.u.] \\ 
         \hline \multicolumn{6}{|c|}{\rule{0pt}{15pt}TD-SC-QED-HF/Pauli-Fierz Hamiltonian} \\ \hline
         sto-3g & -112.35337118& 0.311744 & 0.0001 & 1.26635123 & 0.39477749 \\
6-31g & -113.80630521& 0.311531 &0.0014  &1.26577915  & 0.39432973 \\
6-31g* & -113.86418524& 0.311516 &0.0015  &1.26582131  & 0.39432455  \\
6-31g** & -113.86774654& 0.311512 &0.0016  &1.26580336  & 0.39431348 \\
6-31+g** & -113.87256586& 0.311287 &0.0042  &1.25502739  & 0.39067397 \\
6-31++g** & -113.87273099& 0.310687 &0.0151  &1.15402184  & 0.35853984 \\
aug-cc-pvdz &-113.88439824 & 0.310236 &0.0222  &1.01273656  & 0.31418802 \\
d-aug-cc-pvdz & -113.88476414& 0.309861 &0.0258  &0.88685195  & 0.27480100 \\
d-aug-cc-pvtz  & -113.91293118& 0.310342 &0.0203  &1.03445988  & 0.32103706 \\
         \hline \multicolumn{6}{|c|}{\rule{0pt}{15pt}TD-SC-QED-HF/No Dipole Self-Energy (DSE)} \\ \hline
         sto-3g &-112.35351682 & 0.311743& 0.0001 & 1.26635126    &  0.39477740 \\
6-31g &          -113.80653092 & 0.311530 &0.0014  &1.26577394     & 0.39432692  \\
6-31g* &         -113.86449828&  0.311515& 0.0015 &1.26581618     &  0.39432172  \\
6-31g** &  -113.86806291      & 0.311511 & 0.0016 & 1.26579796 &0.39431052  \\
6-31+g** &       -113.87288665&  0.311278& 0.0044 & 1.25431979    &  0.39044273 \\
6-31++g** &      -113.87305235 & 0.310558 & 0.0176 & 1.12292261    &  0.34873366 \\
aug-cc-pvdz &    -113.88475188&  0.309964& 0.0257 & 0.93366699    &   0.28940407\\
d-aug-cc-pvdz &  -113.88512056&  0.309456& 0.0287 & 0.77936226    & 0.24117891  \\
d-aug-cc-pvtz  & -113.91328959 & 0.310065 & 0.0238 &0.94848608   &   0.29409292  \\
         \hline
    \end{tabular}
    \label{tab: LP nodse sc}
\end{table}
A similar trend is observed for the excitation energy $\omega_{LP}$.
For the transition properties and large basis sets, we notice a significant difference in the photon transition properties $q_{0n}$ and $ip_{0n}$ and an appreciable difference in the oscillator strength $f_l$.
We emphasize that formaldehyde is a small molecule, and these effects are expected to increase with system size and coupling strength $\lambda$.
In addition, the DSE ensures gauge invariance and guarantees the Hamiltonian behaves correctly for large coupling strengths.\cite{schafer2020relevance, rokaj2018light, riso2022molecular}
Since the effect of the DSE on the transition properties can be nonnegligible, we believe including the DSE in the Hamiltonian is preferable, although we do not observe large qualitative changes in the ground state results.

\subsection{Effects of electron-photon correlation on the excited states}

\begin{figure}
    \centering
    \includegraphics[width=\textwidth]{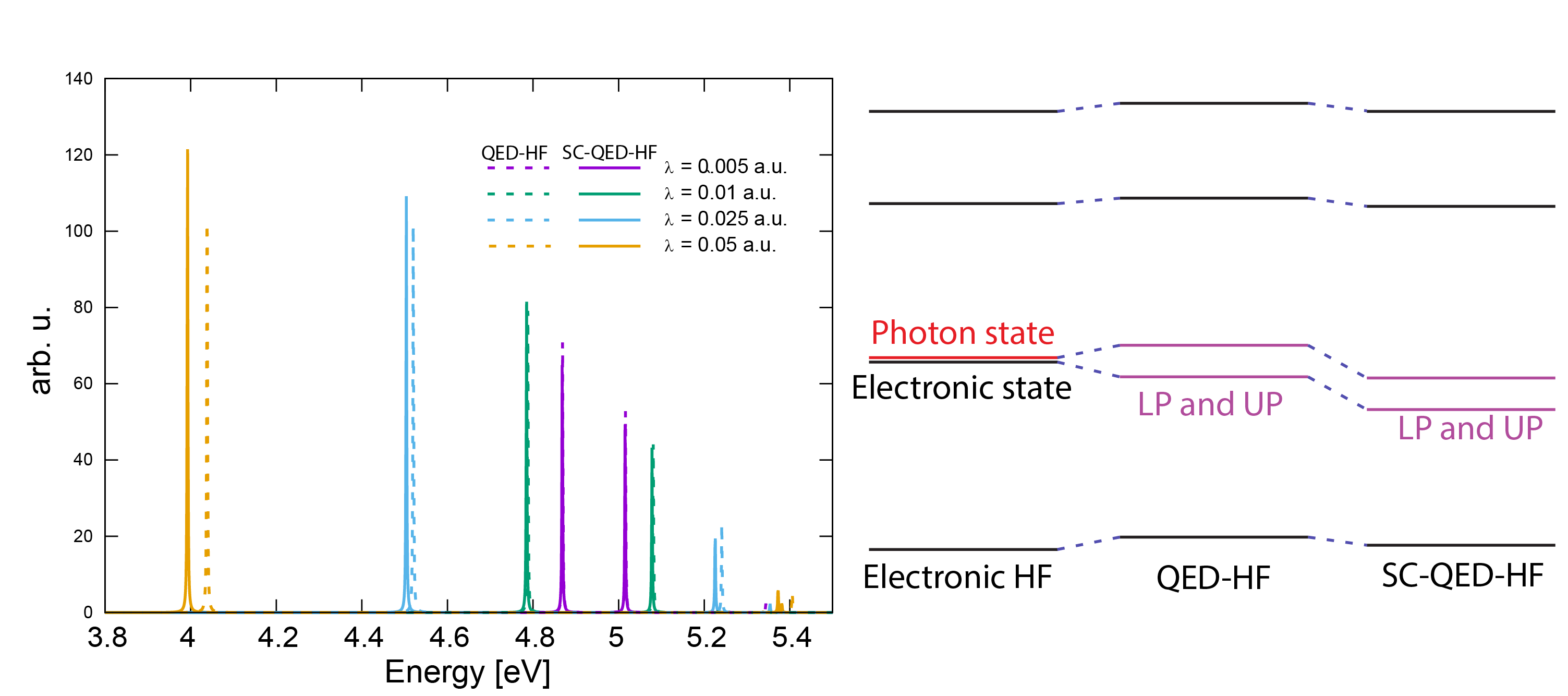}
    \caption{Left: absorption spectrum for the para nitroaniline (PNA) computed at TD-SC-QED-HF (solid lines) and TD-QED-HF (dashed lines) for different coupling strengths $\lambda$ (each polaritonic excitation is endowed with a Lorentzian lineshape).
    The polaritonic excitation energies of TD-SC-QED-HF are lower than the corresponding TD-QED-HF, as shown also in \autoref{tab: lambda exc enenrgy}.
    Right: pictorial representation of the states in the QED-HF and SC-QED-HF models. 
    Compared to the electronic HF results, the QED-HF states have higher energy levels due to the dipole self-energy contribution. 
    Electron-photon correlation then stabilizes the SC-QED-HF states, compared to the mean-field QED-HF theory.
    Since the excited states share a larger photon component, the electron-photon correlation will be larger, thus leading to a more significant stabilization of the excited states compared to the ground state.
    As a result, the excitation energies present a redshift.
    }
    \label{fig: el-phot corr}
\end{figure}

\begin{figure}
    \centering
    \includegraphics[width=\textwidth]{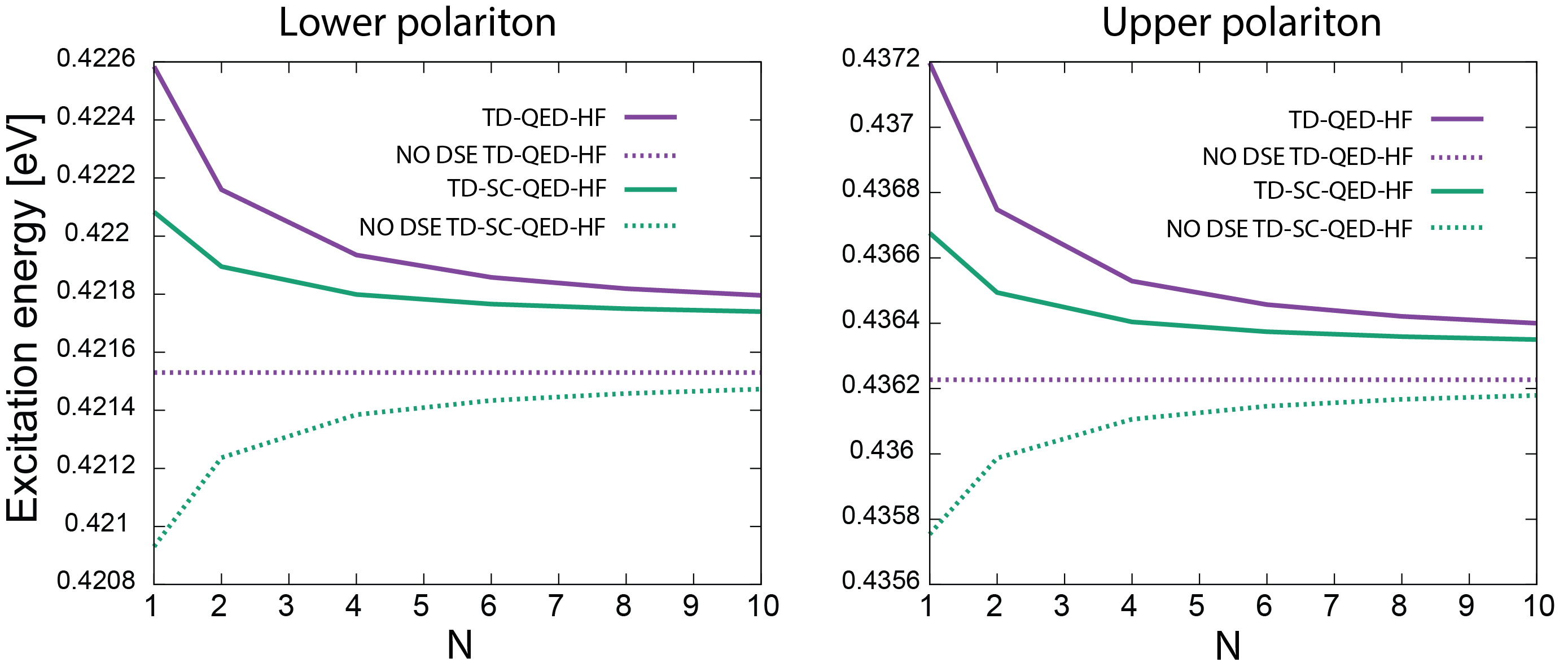}
    \caption{
    Lower polariton (left) and upper polariton (right) excitation energies computed using TD-QED-HF and TD-SC-QED-HF, including (solid lines) and disregarding (dotted lines) the dipole self-energy (DSE).
    We notice that the SC-QED-HF excitations are redshifted compared to the mean field QED-HF energies due to the electron-photon correlation.
    For QED-HF, when the DSE is disregarded, the states depend only on the collective coupling strength $\lambda\sqrt{N}$.
    In the collective regime, the TD-QED-HF and TD-SC-QED-HF energy differences are less pronounced than in the single-molecule simulations, which suggests that the electron-photon correlation depends mainly on the microscopic coupling strength $\lambda$.
    }
    \label{fig: collective comp}
\end{figure}

In \autoref{tab: lambda exc enenrgy}, we report the LP excitation energy $\omega_{LP}$, the length gauge oscillator strength $f_l$, the photon character $\vartheta_{LP}$ in \autoref{eq: photon char param}, and the photon momentum and coordinate transition moments for the p-nitroaniline (PNA) in the aug-cc-pVDZ basis set, computed using the TD-QED-HF and TD-SC-QED-HF models for different coupling strengths $\lambda$.
\begin{table}
    \centering
    \caption{Excitation energy $\omega_{LP}$, length gauge oscillator strength $f_l$, photon character $\vartheta_{LP}$, the transition photon momentum $ip_{0n}$ and coordinate $q_{0n}$ for the lower polariton (LP) of the p-nitroaniline (PNA) computed using the TD-QED-HF and TD-SC-QED-HF models for different coupling strengths $\lambda$ using the aug-cc-pVDZ basis set. 
    The photon frequency is here set resonant to the first bright electronic excitation of PNA (with charge transfer character), and the polarization is along the transition dipole moment (along the C$_2$ axis of PNA).
    The molecular geometry is reported in the Supporting Information.
    }
    \begin{tabular}{|c|c|c|c|c|c|c|}
         \hline $\lambda$ [a.u.] &   E$_\text{GS}$  [a.u.]& $\omega_{LP} $ [a.u.] &   $f_l$  & $\vartheta_{n}$ &    $ip_{0n}$ [a.u.]&   $q_{0n}$ [a.u.] \\ 
         \hline \multicolumn{7}{|c|}{\rule{0pt}{15pt}TD-SC-QED-HF} \\ \hline
          0.005 &  -489.27597426 & 0.178908 & 0.22644 & 0.5312 & 1.22775 & 0.21965 \\
          0.01  &  -489.23850592 & 0.175842 & 0.25831 & 0.5588 & 1.28009 & 0.22509\\
          0.025 &  -489.26686246 & 0.165519 & 0.34288 & 0.6198 & 1.42221 & 0.23540\\
          0.05  &  -489.27483426 & 0.146729 & 0.41351 & 0.6685 & 1.62482 & 0.23841\\
         \hline \multicolumn{7}{|c|}{\rule{0pt}{15pt}TD-QED-HF} \\ \hline
          0.005 &  -489.27583031 & 0.178936 & 0.22418 & 0.5445 & 1.23350 & 0.22071\\
          0.01  &  -489.27425941 & 0.175950 & 0.25341 & 0.5865 & 1.29089 & 0.22713\\
          0.025 &  -489.26330961 & 0.166114 & 0.33003 & 0.6935 & 1.44337 & 0.23976\\
          0.05  &  -489.22479115 & 0.148411 & 0.39369 & 0.8139 & 1.64752 & 0.24451\\
         \hline
    \end{tabular}
    \label{tab: lambda exc enenrgy}
\end{table}
The employed coupling constants correspond to the following quantization volumes: $V \approx $ \qtylist{74.4; 18.6; 3.0; 0.7}{\nanometer\cubed}.
In experimental setups, light-matter strong coupling is achieved via a \textit{collective} coupling, with several ($\sim 10^{5}-10^7$) molecules interacting with the same optical mode.
Some of the employed single-molecule couplings are unrealistic for current experimental devices, and it is still unclear if and to what extent a single-molecule calculation with artificially large coupling can reproduce the effect of collective strong coupling.\cite{castagnola2024collective, castagnola2024realistic, luk2017multiscale, dutta2024thermal, perez2023simulating, horak2024analytic}
Nevertheless, we are here interested in the effect of electron-photon correlation on the wave function parametrization rather than in a careful comparison with experimental results.
The use of large single-molecule coupling is thus not a limitation for this study, and we will later address the effect of the collective coupling.
In \autoref{tab: lambda exc enenrgy}, we notice that the photon character $\vartheta_{LP}$ increases with the coupling strength, following the same trend as the oscillator strength.
This counterintuitive trend (the photon states carry zero oscillator strength) was already emphasized in Ref.\citenum{yang2021quantum} and was explained with the intrusion of higher energy states in the excitation, which compensate for the larger photon character.\cite{yang2021quantum}
Here, we notice that the transition photon coordinate $\braket{\text{GS}|q|\text{LP}}$ and momentum  $\braket{\text{GS}|ip|\text{LP}}$ also follow the same trend.
The increased excitation strength can then be rationalized via \autoref{eq: photo_momenta_conjugate}
\begin{equation}
    \braket{0|\bm{d}\cdot\bm{\lambda}|n}=\frac{\omega_n^2-\omega^2}{\omega}\braket{0|q|n}=i\frac{\omega_n^2-\omega^2}{\omega_n\omega}\braket{0|p|n},\label{eq: rearranged eq rel p-r}
\end{equation}
from which we notice that the transition moment along the polarization direction increases with the transition photon moments and with the Rabi splitting.
In \autoref{tab: lambda exc enenrgy}, we also notice that the TD-SC-QED-HF excitation energies are lower than the TD-QED-HF energies.
Since the two models have the same response parametrization and zero-coupling limit, we can rationalize this result in terms of electron-photon correlation.
As explained in \autoref{sec: theory}, the SC-QED-HF ground state energy is always lower than the QED-HF energy, as can be verified in \autoref{tab: lambda exc enenrgy}.
This is a consequence of the variational optimization of the wave function and is attributed to electron-photon correlation.
Nevertheless, the photon contribution to the ground state is relatively small, even for large couplings.
On the other hand, the excited states share a larger photon component since the electronic excitation is resonant with the cavity frequency.
As a result, the electron-photon correlation is expected to be more relevant for the excited states, which should then be more stabilized than the ground state, as pictorially illustrated in \autoref{fig: el-phot corr}.
Therefore, we expect electron-photon correlation in TD-SC-QED-HF to generally induce a redshift in the excitation energies compared to the TD-QED-HF results.
The frequency redshift for TD-SC-QED-HF also contributes to the asymmetry in the Rabi splitting and, following \autoref{eq: rearranged eq rel p-r}, the LP intensity is generally larger for TD-SC-QED-HF than TD-QED-HF.
The asymmetry of the spectrum also arises from the dipole self-energy, which effectively shifts the molecular electronic excitation, and from the contribution of higher energy states coupled to the cavity photon.
The predicted redshifts of the excitation energies are more relevant for large coupling strengths $\lambda$, while the TD-QED-HF and TD-SC-QED-HF results are more similar for smaller couplings, as expected since both methods converge to the Hartree-Fock excitations (and the one-photon line) for $\lambda\to 0 $.
Since experimental setups rely on a large \textit{collective} coupling, achieved with a relatively small single-molecule coupling $\lambda$ and a large number of molecules coupled to the optical device, it is interesting to study whether such a redshift is also present in the collective regime.
To this end, we focus on a smaller system, the hydrofluoric acid, described using the 6-31++g* basis set, to include more molecules in the simulation.
In \autoref{fig: collective comp}, we report the lower (LP) and upper (UP) polaritonic excitation energies for $N$ fluoridic acid molecules, with coupling strength $\lambda$ such that $\lambda\sqrt{N}=$ \qtylist{0.05}{\atomicunit} for TD-QED-HF and TD-SC-QED-HF.
To focus specifically on the effect of the electron-photon correlation, we also report the results computed by disregarding the DSE, which effectively renormalizes the electronic excitation and thus contributes to an effective cavity detuning.
In \autoref{fig: collective comp}, we notice that the DSE has an appreciable effect on the excited states, even in the collective regime.
We also notice that for QED-HF without the DSE, the upper and lower polariton energies do not change with $N$: the excitation energies depend only on the \textit{collective} coupling strength $\lambda\sqrt{N}$, in contrast to the TD-SC-QED-HF results which show $\lambda$ dispersion even when the DSE is neglected due to the electron-photon correlation.
In \autoref{fig: collective comp}, we see that the TD-SC-QED-HF energies are redshifted compared to the TD-QED-HF excitations also in a collective-coupling setup, even though the differences are less relevant compared to the single-molecule case.
The difference between the TD-QED-HF and TD-SC-QED-HF energies decreases with $N$, both including or disregarding the DSE, which suggests that the electron-photon correlation also depends on the single molecule coupling $\lambda$.
Finally, we notice that the Rabi splitting is asymmetric even when the DSE is not included in the Hamiltonian for both methods in the single molecule and collective regimes (in contrast to what is expected from the two-level Jaynes-Cummings or Tavis-Cummings models).
This is due to the nonperturbative nature of the \textit{ab initio} QED approaches, which implicitly account for all the excited higher-energy states contributing to the polaritons.

\section{Conclusions} \label{sec: conclusions}

In this paper, we have developed and implemented the linear response functions for the recently developed strong coupling quantum electrodynamics Hartree-Fock model (SC-QED-HF).\cite{riso2022molecular}
The SC-QED-HF method is based on the exact ansatz in the infinite coupling limit, and the chosen time-dependent parametrization ensures that TD-SC-QED-HF recovers the exact excited state solutions for $\lambda\to\infty$.
At the same time, the (TD-)SC-QED-HF model converges to (TD-)QED-HF for small couplings.
With the chosen time-dependent parametrization, the TD-QED-HF and TD-SC-QED-HF response equations have the same structure, the difference being only in the Hamiltonian transformation and the optimized molecular orbitals.
We showed that for the developed TD-SC-QED-HF theory, the equivalence relations between the transition moments of photon and molecular observables are fulfilled, similarly to the TD-QED-HF model.\cite{castagnola2024polaritonic, olsen1985linear}
In particular, the equivalence relation between dipole and velocity transition moments is fulfilled in a complete basis set (\autoref{eq: dipole-velocity-form}).
Analogous relations hold for the photon coordinate and momentum (\autoref{eq: dipole-velocity-form photons}), but the relations involving the photonic boson operators hold exactly for any basis set, as demonstrated numerically in \autoref{sec: results} (see the Supporting Information for the analytical proof).
In \autoref{sec: results}, we explored the role of the dipole self-energy and compared the TD-SC-QED-HF results with the TD-QED-HF model.
Since the SC-QED-HF ansatz introduces electron-photon correlation by explicitly mixing the electronic and electromagnetic degrees of freedom, comparing the SC-QED-HF and QED-HF results reveals the effect of electron-photon correlation in the ground and excited states.
Our results suggest that electron-photon correlation induces a redshift in the polaritonic excitations compared to the mean field QED-HF results.
However, our results for a collective-coupling ensemble suggest that electron-photon correlation is also connected to the microscopic light-matter coupling $\lambda$.

Our method provides another step in the development of \textit{ab initio} QED methods based on the consistent SC-QED-HF wave function and provides an additional tool to analyze the effect of light-matter strong coupling on chemical properties.
Since the SC-QED-HF convergence has been recently optimized by using second-order methods\cite{el2024toward}, future works will be devoted to the development of post-HF methodologies and higher-order response functions based on the more consistent SC-QED-HF orbitals.

\begin{acknowledgement}

M.C., R.R.R., Y.E.M. and H.K. acknowledge funding from the European Research Council (ERC) under the European Union’s Horizon 2020 Research and Innovation Programme (grant agreement No. 101020016).
E.R acknowledges funding from the European Research Council (ERC) under the European Union’s Horizon Europe Research and Innovation Programme (Grant n. ERC-StG-2021-101040197 - QED-SPIN).

\end{acknowledgement}

\section*{Data avalability}

The $e^\mathcal{T}$ outputs are available in the following repository: \doi{10.5281/zenodo.14577325}.

\section*{Supporting Information}
The supporting information includes further theoretical details and the molecular geometries employed in \autoref{sec: results}.

\bibliography{bib}
\clearpage
\includepdf[pages=-]{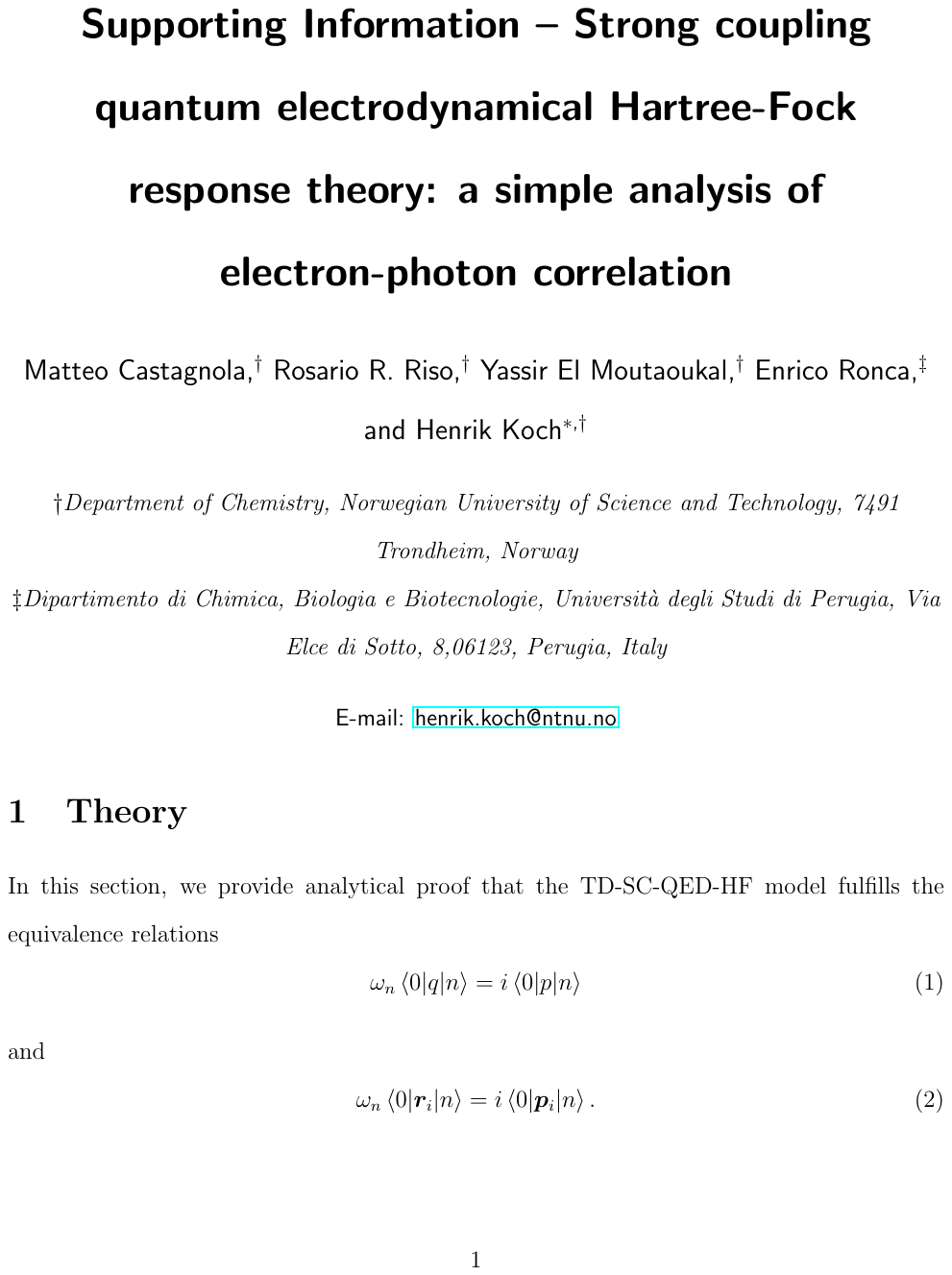}

\end{document}